\documentclass[runningheads]{svmult}

\usepackage{makeidx}   
\usepackage{graphicx}  
\usepackage{subeqnar}  
\usepackage{multicol}  
\usepackage{physprbb}  
\makeindex             

\newcommand{\beq}{\begin{equation}}
\newcommand{\eeq}{\end{equation}}

\def\farcs{\hbox{$.\!\!^{\prime\prime}$}}

\newcommand{\AmS}{{\protect\the\textfont2
  A\kern-.1667em\lower.5ex\hbox{M}\kern-.125emS}}

\begin{document}
\title*{Surface Brightness Fluctuations:\protect\newline
A Case for Extremely Large Telescopes}
\toctitle{Surface Brightness Fluctuations:\protect\newline 
A Case for Extremely Large Telescopes}
\titlerunning{SBF: A Case for Extremely Large Telescopes}
\author{Dimitrios Gouliermis
\and Wolfgang Brandner
\and David Butler
\and Stefan Hippler}
\authorrunning{Dimitrios Gouliermis et al.}
\institute{Max-Planck-Institut f\"{u}r Astronomie, K\"{o}nigstuhl 17, 
D-69117 Heidelberg, Germany}

\maketitle              

\begin{abstract}
The Surface Brightness Fluctuations (SBF) Method for distance
determinations of elliptical galaxies is been modeled in order to
investigate the effect of the Point Spread Function (PSF). We developed a
method to simulate observations of SBF of galaxies having various
properties and located at different distances. We will use this method in
order to test the accuracy on the estimates of the extra-galactic
distances for PSFs representing typical seeing conditions, AO systems and
for future observations with ELTs close to the diffraction limit.
\end{abstract}

\section{The SBF Method}

The SBF Method (Tonry \& Schneider 1988) measures the irreducible mottling in
an early-type galaxy image due to the Poisson fluctuations in the finite
number of stars per pixel. The method has important applications for both
extra-galactic distance (e.g. Jensen et al. 1998) and stellar population
(e.g. Blakeslee et al. 2001) studies. Its advantage is based on the fact that
if we observe the same region of two identical elliptical galaxies at
different distances, the one being twice as far away as the other, the more
distant galaxy would have four times more stars contributing the same average
flux into the pixel as the stars of the closer. The pixel-to-pixel variation
in the flux due to fluctuations in the number of stars then scales inversely
to the distance and so the fluctuations in surface brightness can be used as
a distance indicator.

\begin{figure}[t]
\begin{center}
\includegraphics[height=\textwidth,angle=270]{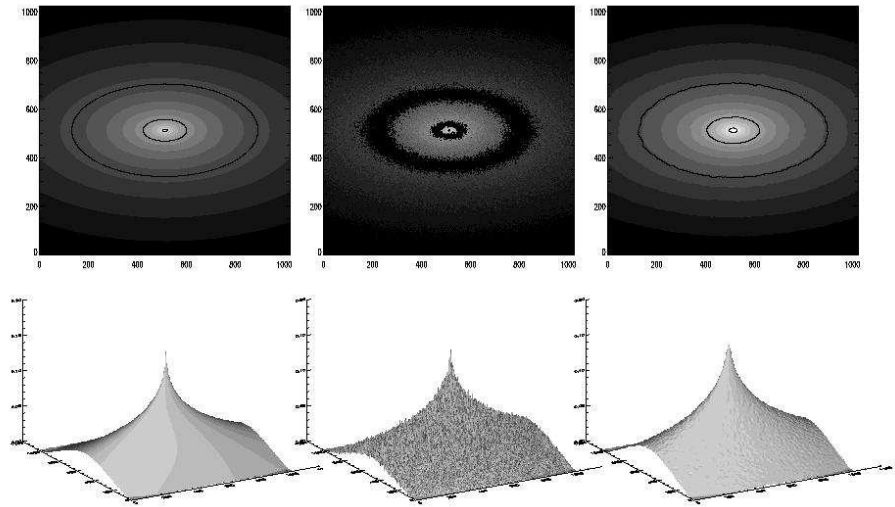}
\end{center}
\caption[]{Grayscale contours (top panel) and 3D brightness distributions
(bottom panel) of a simulated galaxy.  The example shows an E5 galaxy with 
$R_{\rm eff} \simeq 7.5$ kpc located at 20 Mpc following a de Vaucouleurs 
profile. The three steps of the simulations are shown: The model of galaxy's 
brightness profile ({\it right}), its SBF ({\it middle})  and the image 
convolved with a Gaussian PSF of FWHM = 5 px ({\it left}).  A 1024 
$\times$ 1024 pixel CCD array with a pixel scale of 0\farcs 146 was assumed.}
\label{fig1}
\end{figure}

\section{Simulations of Observations on Elliptical Galaxies}

We simulate surface brightness profiles and the SBF of
elliptical galaxies {\em at distances of our choice}. We then convolve them
with various kinds of PSF in order to simulate their observations. The PSF is
considered for three cases: {\em (}$\alpha${\em )} seeing-limited
observations, {\em (}$\beta${\em )} observations with nowadays available
{\em Adaptive Optics} (AO) systems, and {\em (}$\gamma${\em )} with {\em 
Multi-Conjugate Adaptive Optics} (MCAO) systems, which are to be used for 
ELTs. Thus, the first part of our method consists of three steps (see Fig. 
\ref{fig1}):
\begin{enumerate}
\item \underline{Modeling of the Surface Brightness Profiles of typical 
elliptical galaxies.}\\ 
The simulated surface brightness profiles are assumed to follow Sersic
($r^{1/n}$) law. A profile for $n = 4$ (de Vaucouleurs law) is given as:
\beq
\mu(r)=\mu_{\rm e}+8.3268~\left[{\left(\frac{r}{r_{\rm 
e}}\right)}^{1/4}-1\right]{\rm and}~\mu_{\rm e} = 5~\log{(r_{\rm e})} 
+ m + zp + c
\eeq
where $\mu_{\rm e}$ is the surface brightness of the galaxy at its effective
radius $r_{\rm e}$, while $m$ is its apparent magnitude. Among the input 
parameters for this step is the distance of the galaxy, so we decide in 
advance how far away the galaxy is located. We are also able to define 
the wave-band at which the galaxy is being observed.
\item \underline{Modeling the surface brightness fluctuations of the 
galaxies.}\\ 
The reproduction of the SBF of the simulated galaxies is done by the 
introduction of Poisson noise to the surface stellar density of every pixel 
of the model image assuming a luminosity function for the stellar 
population of each galaxy.
\item \underline{Convolution with the PSF.}\\
This is the final step for the construction of simulated observations of 
elliptical galaxies on arrays with size and resolution of our choice.
\end{enumerate}


\begin{figure}[t]
\begin{center}
\includegraphics[width=\textwidth]{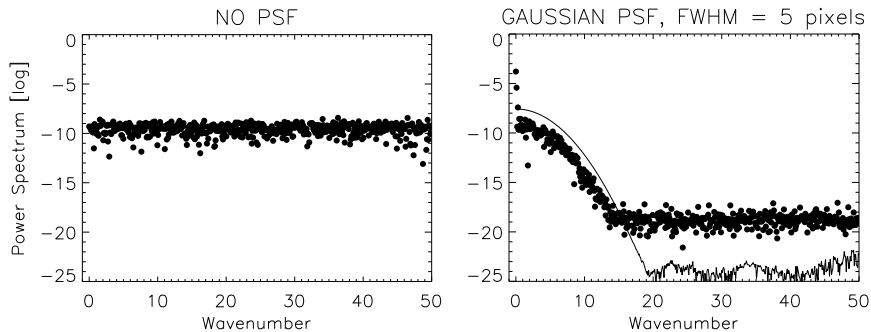}
\end{center}
\caption[]{ Power spectrum of a simulated E5 galaxy with $R_{\rm eff} = 10$
kpc, located at 35 Mpc. {\it Left}: The power spectrum of the image when
there is no PSF effect ({\it Left}) and for seeing 0\farcs73 ({\it Right}).
In the later the line shows the power spectrum of the PSF.}
\label{fig2}
\end{figure}

\section{Measurements of the SBF of the Simulated Galaxies}

The second part of our method includes the measurements of the SBF of the
simulated galaxies, the computation of their SBF magnitude and in consequence
the estimation of the distance of the galaxies. In order to measure the 
SBF of an observed galaxy the power spectrum of the reduced image must be 
computed. It is given as the linear combination of the fluctuation power 
$P_{0}$ times the power spectrum of the PSF $E_{\rm psf}$ and a white noise 
component, $P_{\rm w}$:
\beq
P(k) = P_{0}E_{\rm psf}+P_{\rm w}
\eeq
In consequence {\it the power spectrum of the PSF defines the one of the 
image of the galaxy itself}. This is demonstrated in Fig. \ref{fig2}, where 
we show the power spectrum of simulated observations on an elliptical galaxy 
without any PSF at all (left) and with seeing limited conditions (right). 

\section{Conclusion}
We present a model of the SBF method for distance estimations. Our aim is to
use it for the investigation of hypothetical SBF observations with ELTs.
Specifically, the measurements of the SBF amplitude of simulated observations
of elliptical galaxies with various properties can lead to estimations of the
distances of the galaxies with the use of available SBF calibrations (e.g.
Tonry et al.  2001). The comparison between the input distance that was
initialy selected for every galaxy and the estimated value will allow us to
define the accuracy of the SBF method with the use of various telescopes and
under various observational conditions. Thus we will be able to study the
accuracy and the limitations of SBF observations with future ELTs.


%

\end{document}